# Helium under high pressure: A comparative study of all-electron and pseudopotential methods within density functional theory

W. Xiao, Z. X. Tian and W. T. Geng[a]

*School of Materials Science & Engineering, University of Science & Technology Beijing, Beijing 100083, China,*

We have calculated the ground state electronic structure of He under pressure from 0 to 1500 GPa using both all-electron full-potential and pseudopotential methods based on the density functional theory (DFT). We find that throughout this pressure range, pseudopotentials yield essentially the same energy-volume curve for all of *bcc*, *fcc*, and *hcp* configurations as does the full-potential method, a strong indication that pseudopotential approximation works well for He both as the common element in some giant planets and as detrimental impurities in fusion reactor materials. The *hcp* lattice is always the most stable structure and *bcc* the least stable one. Since the energy preference of *hcp* over *fcc* and *bcc* is within 0.01 eV below 100 GPa and about 0.1 eV at 1500 GPa, on the same order of the error bar in local or semi-local density approximations in DFT, phase transitions can only be discussed with more precise description of electron correlation in Quantum Monte Carlo or DFT-based *GW* methods.

---

[a] To whom correspondence should be addressed. E-mail: geng@ustb.edu.cn

# I. INTRODUCTION

The lightest inert gas, helium, rarely occurs in our everyday life, but appears as one common element in the field of astrophysics.[1] In universe, hydrogen, helium or a mixture thereof are the major component of the interiors of Jupiter, Saturn and gaseous extrasolar planets, as well as white dwarf atmospheres.[2-6] Therefore, understanding the behavior of helium at high pressure and high temperature is of great importance in our attempt to comprehend the formation, evolution and fate of many extraterrestrial bodies. Over the last few decades, many experiment and simulation techniques have been applied to explore its equation of states (EOS) under these extreme conditions. Early attempts to describe dense He using simple empirical pair potentials with harmonic approximation returned poor agreement with experiments below 100 Mpa, yet a good comparison at 15 GPa.[7] This is a strong indication that a uniform approximation for the He-He potential cover both core and peripheral region is hard to define. With Linear-Muffin-Tin-Orbitals (LMTO) computation based on DFT, Young *et al.*[8] explored the EOS of He under pressure as high as 25 TPa, a situation similar to the interior of giant planets. Metalization of He was predicted to occur at 11.2 TPa; yet no experimental data were available then. Mao *et al.*[9] demonstrated with synchrotron radiation through single-crystal x-ray diffraction measurements that the structure of the solid helium is hexagonal close packed from 15.6 to 23.3 GPa at 300 K. And Hugoniot data for liquid helium has been secured by shock experiments in the 100 GPa pressure range,[10] which tested EOS models over a broader range and separated the effect of temperature and density on the EOS.

Inspired by experimental advancement, precise DFT-based first-principles methods have recently been intensively applied to the investigation of dense helium at high pressures. Combining path integral Monte Carlo (PIMC) and density molecular dynamics (DFT-MD), Militzer studied the properties of hot dense He and predicted the maximum compression ratio of He/D.[11, 12] In a more recent work, Khairallah and Militzer examined in detail the effect of approximation in exchange-correlation on the electronic structure of highly densed He.[13] Interestingly, they found that the generalized gradient approximation (GGA) in DFT underestimate both density and pressure for the non-metal to metal transition, compared with the diffusion quantum Monte Carlo (DMC) method, which is viewed as truly exact. The *GW* band gap corrections, nonetheless, were in good agreement with DMC calculations. Furthermore, these authors scrutinized the zero point motion of the nuclei, and found no noticeable effect on the metallization within the accuracy of the calculation.

In searching for the possible error introduced by approximations made in first-principles computational methods, the only stone left unturned is the usage of pseudopotentials in previous studies. Since He has only two 1*s* electrons which are always treated as valence electrons, there are no core electrons to freeze when pseudopotentials are employed. However, the deviation of pseudopotential from the true one in the core region may still incur some error to the calculated states of electrons, especially under high pressure conditions.

In parallel to the study of He in giant planets, much attention has also been paid in the last decade on the investigation of the behavior of He in metals, which presents a great challenge in design of structural materials used in fusion reactors.[14] An appreciable content of H and He will be unavoidably introduced into metals under transmutation reactions.[15] The insoluble He will lead to hardening and shortened fatigue life by acting as an obstacle to the movement of dislocations at low temperatures. Also, it will result in significant degradation of the tensile, creep and fatigue properties at high temperatures.[16] With high dose, He atoms tend to form platelets and bubbles very easily because He-He bound strongly in metals. The maximum pressure of a He bubble is estimated to be about 20% of the shear modulus of the host material.[14, 17] For instance, it can be as high as dozens of GPa in low activation matensitic steel, which are viewed as promising candidates for structural materials used in fusion reactors. Since the volume of a single atom is not well defined, it is hard to discuss the pressure exerted on a single He atom. The solution energy of an interstitial He in Fe is 4.37 eV,[18] equal to one in a bubble with a pressure of several hundreds of GPa. This means that He atoms in metals, either in separated or accumulated positions, are under pressure. As for the He migration energy in W, there is a huge discrepancy between experiment (0.24 eV) and pseudopotential calculations (0.06 eV).[19] Becquart and Domain proposed that the measurements might be done for He pairs, rather than single He because pairs are very easy to form. Although the pseudopotential approximation alone is not expected to account for such a significant disagreement, we are still very curious about how much an error this approximation introduces.

Motivated by these open questions, we have performed the first-principles DFT calculations on both gaseous He and that in Fe, using both the projector augmented wave (PAW) method[20, 21] implemented in the Vienna *ab initio* Simulation Package (VASP) and the all-electron full potential linearized augmented plane waves (FLAPW) method[22, 23] implemented in Wien2k.

## II. MODEL AND COMPUTATION

We placed the He atoms on three kinds of lattices, face-centered cubic (*fcc*), body-centered cubic (*bcc*) and Hexagonal close-packed (*hcp*), and calculated the cohesive energy as a function of volume (per atom) for each system. The step of volume change was 5.0 a.u.$^3$ for all lattices, down from around the equilibrium volume 120.0 a.u.$^3$ (10.7cm$^3$/mol).[24]

For both VASP and Wien2k calculations, the exchange-correlation of electrons were depicted with GGA in the Perdew-Burke-Ernzerhof (PBE) form[25]. The 1*s* electrons of He were treated semirelativistically, i.e., without spin-orbit coupling. The Brillouin-zone integration was performed within Monkhorst-Pack scheme using a (12×12×12) *k*-mesh for *fcc* and *bcc* and a (12×12×6) *k*-mesh for *hcp*. Tests on denser *k*-meshes show that the convergence of cohesive energy on *k*-mesh is well within 0.001 eV.

For VASP calculations, the energy cutoff for the plane wave basis set was 600 eV, to ensure a precise description of the low-lying 1*s* states of He. The radius of the core within which pseudopotential was employed was 1.1 a.u.. In the FLAPW method, no shape approximations are made to the charge densities, potentials, and matrix elements. For Wien2k calculations, an energy cutoff of 64 Ry was employed for the augmented plane-wave basis to describe the wave functions in the interstitial region, and a 400 Ry cutoff was used for the star functions depicting the charge density and potential. The muffin-tin radius of He is always fixed at 1.0 a.u.. Within the muffin-tin spheres, lattice harmonics with angular-momentum $l$ up to 10 were adopted to expand the charge density, potential, and wave functions. In both sets of calculations, the convergence of cohesive energy on plane wave basis set is well within 0.001 eV.

As for He in Fe, we chose a (3×3×3) *bcc* Fe supercell with 54 atoms in both VASP and Wien2k calculations. Previous study[18] showed such a cell is large enough to minimize the boundary effect. The Brillouin-zone integration was performed within Monkhorst-Pack scheme using a (3×3×3) *k*-mesh in both calculations. For all the systems, spin-polarization was permitted.

For VASP calculations, the lattice constant of the *bcc* Fe was calculated to be 2.83 Å, in good agreement with the previous *ab initio* calculations. The 3*d* and 4*s* electrons of Fe were treated as valence electrons. In the presence of He, the lattice relaxation was continued until the forces on all the atoms were converged to less than $10^{-3}$ eV Å$^{-1}$. The energy cutoff for the plane wave basis set was 480 eV. In the Wien2k method, the calculated lattice constant for the bulk bcc Fe is 2.84 Å. Energy cutoffs of 28 Ry and 196 Ry were employed for plane-wave bases and star functions to describe the wave functions and the charge density in the interstitial region, respectively. Muffin-tin radii for Fe and He atoms were chosen as 2.0 and 1.1 a.u., respectively. Within the muffin-tin spheres, lattice harmonics with angular momentum $l$ up to 10 were adopted. Lattice relaxation was continued until the forces on all the atoms were converged to less than $3\times10^{-3}$ eV Å$^{-1}$.

## III. NUMERICAL RESULTS

In Fig. 1, we display the calculated cohesive energy of He as a function of volume in *fcc*, *bcc* and *hcp* alignments using VASP and Wien2k. As mentioned above, the variation in volume has a step of 5 a.u.$^3$. Two more points below 10 a.u.$^3$, i.e., 8.0 and 9.0 a.u.$^3$ were also calculated. Note that in *bcc* structure, if one He atom has a volume of 8.0 a.u.$^3$ its contacting radius, that is, one half of the nearest-neighbor distance is about 1.1 a.u., equal to the radius of the core within which pseudopotential was employed.

**Fig. 1 (Color online) The calculated cohesive energy of He as a function of volume in *fcc*, *bcc* and *hcp* alignments using VASP (a) and Wien2k (b). Panel (c) shows the comparison of VASP and Wien2k results for *hcp*. Each energy here is in reference to the state at the volume 120.0 a.u.$^3$ derived from the same method. The step of volume change was 5.0 a.u.$^3$ for all lattices, down from the equilibrium volume 120.0 a.u.$^3$ Two more points below 10 a.u.$^3$, i.e., 8.0 and 9.0 a.u.$^3$ were also calculated.**

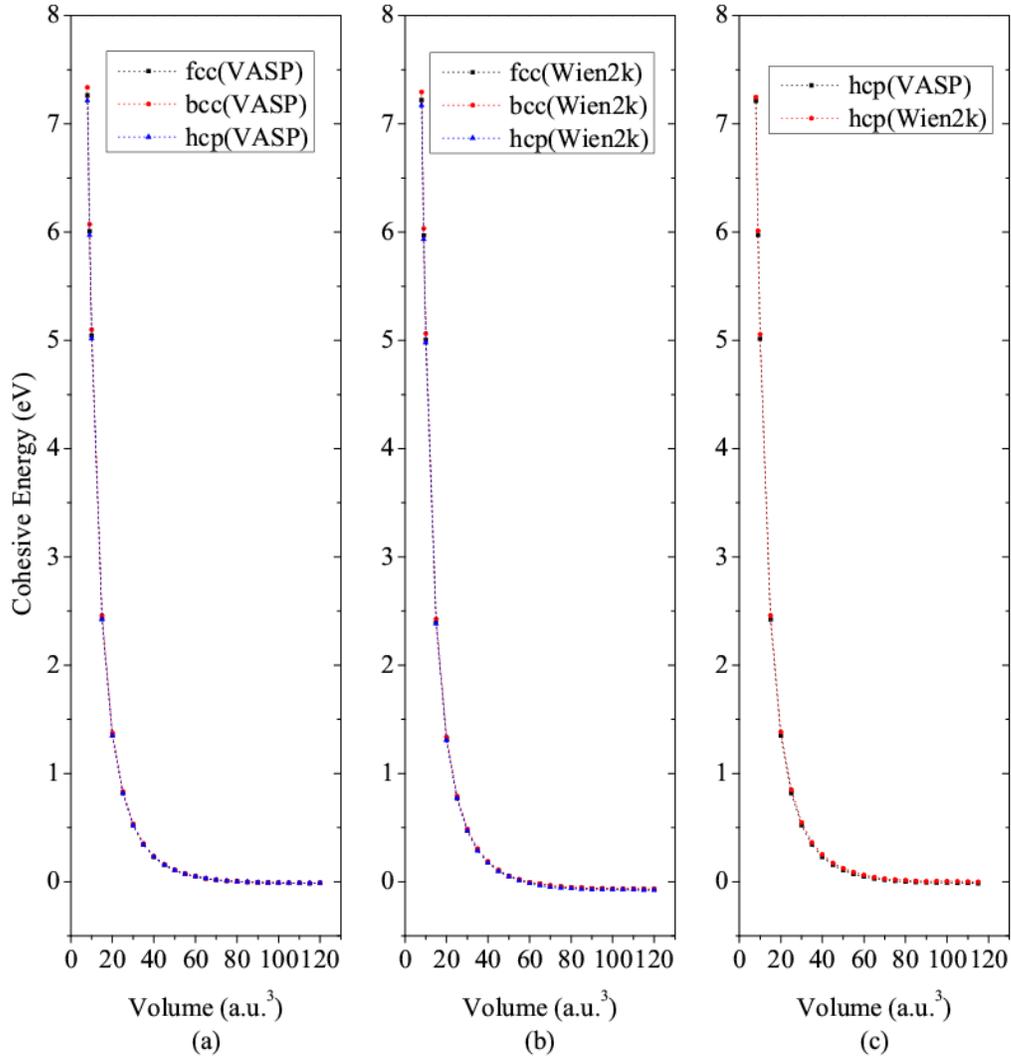

A similarity between panels (a) and (b) is that the cohesive energy of condensed He depends almost only on volume and symmetry plays a very marginal role, as is evident that the three curves in each panel are hard to separate. These numerical results are in accordance with the fact that He is extremely inert and thus very difficult to solidify. But we do find that the *hcp* lattice is always the most stable structure and *bcc* the least stable one. At high pressure, this is different from the results of Young et al.'s early nonrelativistic LMTO calculations with $X$-$\alpha$ potential, which predicted phase transitions.[8] At low pressure, our result agrees well with Loubeyre et al.'s[24] experiment that there is no volume difference between *fcc* and *hcp* phase below 20 GPa. Notice that the energy preference of *hcp* over *fcc* and *bcc* is only about 0.01 eV below 100 GPa and 0.1 eV at 1500 GPa, on the same order of the error bar in local or semi-local density approximations in DFT. We have compared the most commonly used GGA functional, PW91[26] and PBE[25] on the predicted cohesive energy of He. The results are presented in Table I. Therefore, we shall refrain ourselves from discussing phase transitions, which are better discussed with more precise description of electron correlation, such as in Quantum Monte Carlo or DFT-based *GW* methods. In panel (c), it can be seen clearly that the cohesive energies yielded by VASP and Wien2k are very close to each other, a strong indication the influence of pseudopotentials is less significant than exchange-correlation functionals.

**TABLE I. The cohesive energy (eV) for *fcc* He at volumes of, 8, 30 and 120 a.u.$^3$ per atom given by VASP, using two GGA functionals, PW91 and PBE.**

| Volume | 8 a.u.$^3$ | 30 a.u.$^3$ | 120 a.u.$^3$ |
|---|---|---|---|
| PW91 | 7.21 | 0.50 | -0.03 |
| PBE  | 7.26 | 0.52 | -0.01 |

The calculated equilibrium volumes for *fcc*, *bcc* and *hcp* phases using VASP are 117 (*a*=4.11a.u.), 119 (*a*=3.28a.u.) and 118 a.u.$^3$ (*a*=2.91a.u. *a/c*=0.61), respectively. Since the energy-volume curves are very flat around the equilibrium point, an error of 0.01 eV in the cohesive energy will introduce an error in the equilibrium volumes on the order of

a.u.³. By comparison, no equilibrium point is found in Wien2k calculations for the energy becomes completely flat when the volume per a He atom is larger than 100 a.u.³.

**Fig. 2 (Color online) The calculated pressure as a function of volume for one condensed He atom from VASP and Wien2k, respectively.**

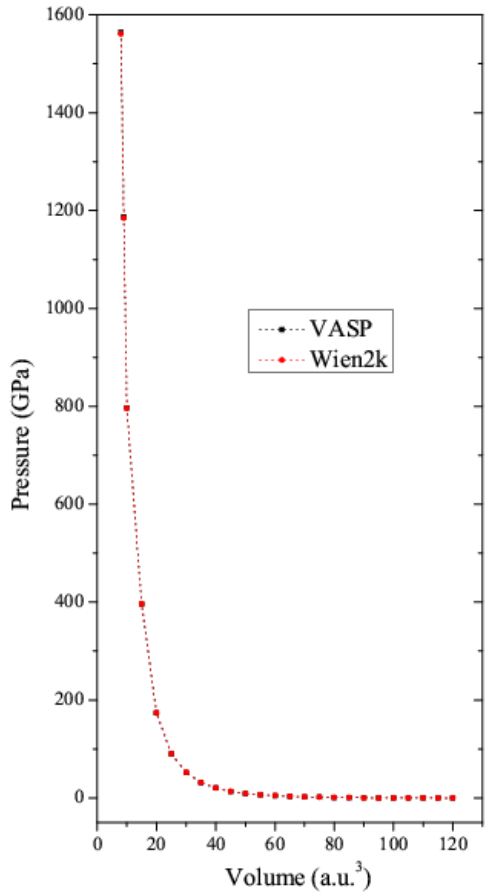

We plot in Fig. 2 the pressure-volume relationship by performing the first derivatives of the energy-volume function. We find when the volume of He is compressed to 25 a.u.³/atom, the pressure is still less than 100 GPa, a pressure spans the He bubbles in metals. When compressed to 10 a.u.³, the pressure of He will dramatically soar to 1000 GPa, attaining the pressure relevant in astrophysics. Again, we find that pseudo-potential method yields essentially the same pressure-volume function for helium as the full potential approach in a wide pressure span.

Helium atoms are always introduced as defects in structural material used in fusion reactors. In this field, many simulations have been employed by pseudo-potential method. In *bcc* Fe, He prefers to occupy the tetrahedral site with a solution energy of 4.37 eV.[18] Such an energy corresponds a pressure of about 500 GPa in Fig. 2. We thus expect that VASP and Wien2k shall give similar electronic states for He in Fe. To verify if this is the case, we have carried out two sets of calculations on the migration energy of an interstitial He in *bcc* Fe using both methods. With Wien2k, we calculated only the energies of the initial (most stable) and transitional (most unstable) configurations in the He diffusion path as shown in [Ref. 18, Fig. 2].[18] We find that VASP gives a diffusion barrier of 0.06 eV, the same as Fu and Willaime's using a different pseudopotential DFT code. The more precise all-electron fullpotential method, Wien2k, on the other hand, predicted it to be 0.07 eV. Our comparative study thus demonstrate unambiguously that pseudopotentials work well for the determination of diffsion of He in Fe, and presumably other metals.

## IV. DISCUSSION

To get more insight into subtle difference between VASP and Wien2k, we have examined carefully the eigenvalues of He. In order to compare more meaningful, we choose the same *k* mesh for the Brillouin-zone integration. We listed in Table II the energy levels determined by both methods. Each value is in reference to the lowest level for that volume, i.e. the first eigenvalue at *k* point (0, 0, 0) is always set to zero. We can see that VASP and Wien2k give the same energy states for helium at the atomic volume of 120 a.u.$^3$; whereas at a high pressure of about 1500 GPa (8 a.u.$^3$) there is a negligible 0.02 eV difference (~ 0.1%). As pressure goes up, electrons from adjacent He atoms interact strongly with each other that greatly enhances the band dispersion of bonding and anti-bonding states.

**TABLE II. The calculated energy levels (eV) for the occupied bands of *fcc* He at atomic volume of 8.0 and 120.0 a.u.$^3$. Data for three *k* points are listed here. Each energy is in reference to the lowest energy state.**

|  |  | volume = 8 a.u.$^3$ | | Volume = 120 a.u.$^3$ | |
|---|---|---|---|---|---|
| *k* points | level | VASP | Wien2k | VASP | Wien2k |
| Γ (0, 0, 0) | 1 | 0 | 0 | 0 | 0 |
|  | 2-4 | 37.52 | 37.50 | 1.01 | 1.01 |
| X (0, 0.5, 0.5) | 1-4 | 23.27 | 23.26 | 0.77 | 0.77 |
| L (0.5, 0.5, 0.5) | 1-4 | 27.86 | 27.84 | 0.79 | 0.79 |

Finally, we plot DOS for *fcc* He at three pressures, which is about 0, 50 and 1500 GPa respectively. At equilibrium point, the occupied DOS has one sharp peak right beneath the Fermi level. The bandgap is about 14.4 eV. As the pressure goes up to 50 GPa, the peak was smeared down to about one quarter of its origin height and the broadening was extended to 10 eV. Under 1500 GPa, the peak was smeared down further and the broadening was continuously stretched up to 45 eV. At the same time, the bandgap decreased slowly, which is consistent with the prediction that the band gap will close at the pressure of 25700 GPa. [Ref. 13]

**Fig. 3 (Color online) The calculated density of states (DOS) using VASP for *fcc* He at volumes of, 8, 30 and 120 a.u.$^3$, respectively. The Fermi energy was set to 0 eV.**

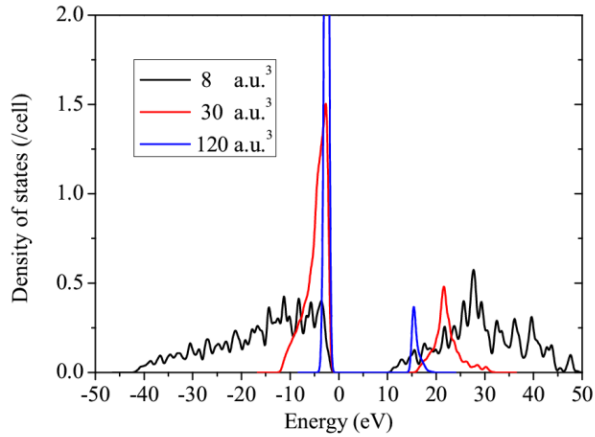

## V. SUMMARY

In conclusion, our first-principles calculations on the ground state electronic structure of He under pressure from 0 to 1500 GPa using both all-electron full-potential and pseudopotential methods based on the density functional theory (DFT) demonstrate that from low to high pressure, pseudopotentials give essentially the same energy-volume curve for all of *bcc*, *fcc*, and *hcp* configurations as does the full-potential method, indicating that pseudopotential approximation works well for He both as the common element in the interior of some giant planets and as detrimental impurities in fusion reactor materials. As a consequence, we can employ with confidence the pseudopotentials to save computational effort.

**ACKNOWLEDGMENTS**

The work was supported by the NSFC (Grant No. 50971029) and MOST (Grant No. 2009GB109004) of China. The calculations were performed on the Quantum Materials Simulator of USTB.